# High-Performance Self-Powered Photoelectrochemical Detection Using Scalable InGaN/GaN Nanowire Arrays

Kishan Lal Kumawat[1], Md. Afjalur Rahman[1], Nirmal Anand[1], Dipon Kumar Ghosh[1], Christy Giji Jenson[1], Md. Moinul Islam[1], Samuel Olakunle Adigbo, Sheik Munim Hussain[2], Md Zunaid Baten[1], and Sharif Md. Sadaf[1]*

[1]Centre Energie, Matériaux et Télécommunications, Institut national de la recherche scientifique (INRS), Varennes, Québec J3X 1P7, Canada

[2]Department of Electrical and Electronic Engineering, Bangladesh University of Engineering and Technology, Dhaka, Bangladesh

* E-mail: sharif.sadaf@inrs.ca

**Abstract**

Photoelectrochemical photodetectors (PEC-PDs) are promising owing to their simple, low-cost fabrication, self-powered operation, high photoresponse, and environmental sensitivity. In this work, we report for the first time the self-powered PEC photodetection characteristics of nanowire (NW)-based green-emitting InGaN/GaN multiple quantum well (MQW) PEC-PDs, fabricated via a scalable top-down approach. The nanowire geometry significantly enhances the photoresponse compared to planar InGaN/GaN photoelectrodes, attributed to the increased solid–liquid interfacial area and improved carrier transport. A maximum photocurrent density of 16 mA/cm² (twice that of the planar structure) was recorded under 365 nm illumination at 43.5 mW/cm² without external bias. The device exhibits strong UV sensitivity with a peak at 365 nm and an extended response into the visible region. Notably, a high photoresponsivity of ~330 mA/W was achieved at a lower illumination intensity of 0.7 mW/cm². Furthermore, the photodetector demonstrates fast, stable, and reproducible performance across varying biases and illumination conditions. These results suggest that InGaN/GaN MQW nanowire-based PEC photodetectors hold strong promise for scalable, efficient, and stable self-powered optoelectronic applications.

**Introduction**

In the era of the Internet of Things (IoT), there is a growing demand for highly sensitive, multi-functional, ultrafast and small-sized sensors /devices for integration into the global positioning system (GPS) network for high-performance communication systems [1-3]. Photodetectors are sensors that can convert optical signals into electrical signals by measuring the resistance change in the device upon light irradiation [4]. Depending on the wavelength detection range of optical signals in the electromagnetic spectrum, photodetectors have promising potential applications in various fields which includes space/optical communication, imaging technology, flame detection, astronomical exploration, artificial vision, environmental monitoring, wearable sensors/smart electronics and biochemical analysis [5-8]. Considering different types of solid-state photodetectors, such as photoconductors, phototransistors, photomultiplier tubes (PMTs), PEC-PDs have gained much attention due to their outstanding performance, including higher photocurrent density, fast photoresponse, easy fabrication process, and ability to operate without external bias energy [9-10]. Unlike the conventional solid-state PDs, PEC-PDs feature a unique solid-liquid interaction for photocarrier generation and transport. The use of liquid electrolyte and involvement of redox reactions provide additional flexibility to optimize the photodetection results and to obtain special features that are otherwise not possible with solid-state photodetectors [11]. The performance of PEC photodetectors is largely governed by the dynamics of photoengenerated charge carriers and ion transport in the electrolyte. By using different light wavelengths irradiation and intensities, the charge carrier dynamics and chemical reactions can be regulated to get a tunable photoresponse for multifunctional optoelectronic devices [6]. However, simultaneously achieving high responsivity and stability, especially under harsh conditions of highly acidic and basic electrolytes, remains challenging for PEC photodetectors [12]. Moreover, it is challenging to achieve miniaturized next-generation smart photodetection with low-power consumption and/or self-powered operation, cost-effectiveness and multifunctionality [13]. To this end, various materials have been recently explored for PEC PDs, including 2D layered metal dichalcogenides such as $MoS_2$, $WSe_2$, $SnSe_2$ and others, metal oxides: ZnO, $TiO_2$, $WO_3$, $SnO_2$, III-nitrides: GaN, InN, AlN, and related alloys [14-20]. Nevertheless, metal oxides and other traditional classes of materials have limitations such as wide band gaps, which limit the detection bandwidth, unsuitable band edge positions for water redox reactions and poor photostability due to oxygen defects.

Among the materials classes recently studied for PEC PD, GaN and InGaN show tremendous potential for the PEC photodetection due to their exceptional properties, such as wide, tunable band gaps (from 3.4 eV to 0.7 eV), high absorption coefficient, high charge carrier mobility, and excellent photostability [21]. While designing a PEC-PDs, both the photoconversion performance of materials and the efficient interfacial redox reactions are critical. Conventional planar heterostructures based PEC devices suffer from lower photoresponsivity due to lower catalytically active surface sites, longer carrier diffusion length, and lower optical absorption of light [22]. Alternatively, NW structure PEC-PDs offers additional advantages such as enhanced surface to volume ratio which increases the available active sites for the chemical redox reactions, flexibility to tune the device parameter by varying the aspect ratios of NWs, enhanced charge carrier separations and transportations due to reduced diffusion path length from micrometer to nanometer scale, and improved crystallinity [22, 12]. Additionally, the anti-reflection properties of NWs enhance light absorption efficiency by promoting light trapping through multiple scattering and offer better control over device parameters [23]. Previous reports have focused on the bottom-up spontaneous III-nitrides NWs for LEDs, solar cells and PDs applications [24-32]. A recent report demonstrated green color InGaN/GaN top-down nanowire array LEDs for near-eye display applications, fabricated by e-beam lithography technique [33]. However, to date, there has been no report on the scalable top-down nanowire fabrication by the silica nanosphere fabrication technique for PEC-PD applications.

In general, InGaN/GaN MQW are commercially utilized in solid-state lighting applications; also such MQW structures have demonstrated significant performance as photoelectrodes for PEC water splitting [34]. However, their use in PEC UV-visible photodetectors is still largely unexplored. Here, we have demonstrated for the first time the self-powered PEC photodetection performance of InGaN/GaN multiple quantum wells (MQWs) NWs photodetectors having an In composition of 25%. In this study, we employed a scalable top-down approach silica nanosphere lithography technique to fabricate NWs PD of MOCVD grown InGaN/GaN MQWs heterostructure. The PEC photodetector exhibited a high responsivity of 330 mA/cm$^2$ under self-powered mode with irradiation of light intensity of 0.7 mW/cm$^2$, which is the highest as reported yet among the III-nitride based PEC-PDs. The high figures of merit are attributed to the better separation and transportation of photogenerated charge carriers due to MQW heterostructures and the induced built-in electric potential.

Moreover, InGaN/GaN MQW nanowires show better photocurrent density than that of planar heterostructure-based devices due to a larger area of solid-liquid reaction interface and directional charge carrier transmission provided by the NWs. The photocurrent density increased almost double as compared to the planar device without any external power source under 365 nm light wavelength, with the intensity of 43.5 mW/cm$^2$. Furthermore, the InGaN/GaN NW PEC-PD device exhibits excellent photostability and a fast, repeatable photoresponse under varying applied biases and light intensities, highlighting its potential for scalable, self-powered, and efficient PEC photodetection applications.

**Results and discussion**

The structure of the InGaN/GaN MQWs with ~25% indium content was grown heteroepitaxially on (0001) sapphire substrate under optimized growth conditions by metal organic chemical vapor deposition (MOCVD). The epitaxial structure consists of a 20nm GaN buffer layer, 1.75 μm thick n-type GaN layer, 10 pairs of InGaN/GaN (3.1 nm)/(15 nm) MQWs, a 20 nm thick p-type $Al_{0.2}Ga_{0.8}N$ electron blocking layer, and a 70 nm thick Mg-doped p-GaN cap layer. Figure 1(a) shows the schematic illustration of the NW of the InGaN/GaN MQW heterostructure. A top-down etched silica nanosphere lithography technique was employed for the NW structure formation. The surface morphology of the etched nanowires are determined by SEM and shown in Figure 1(b). The average diameter and height of the NW was around 330 nm and 900 nm.

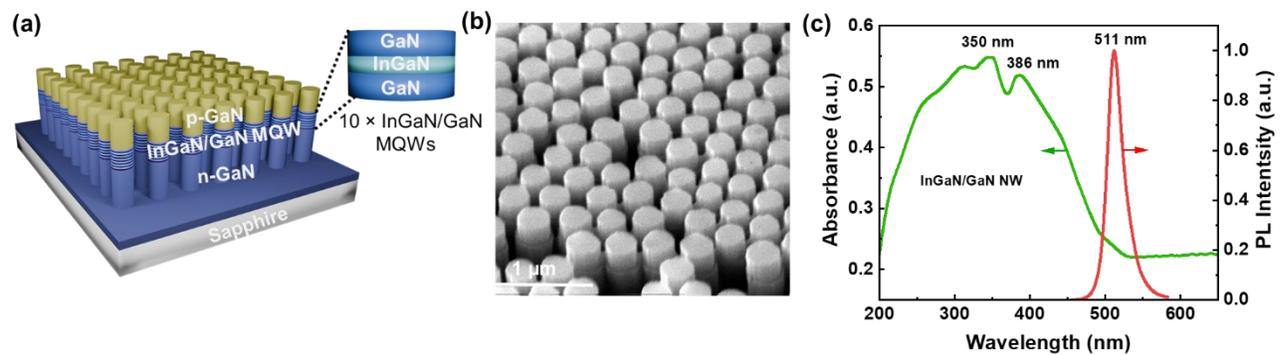

**Figure 1(a and b):** Schematic of NW structured InGaN/GaN heterostructures fabricated via top-down etch silica nanosphere patterning method coupled with dry and wet etching and the corresponding SEM image of NWs. **(b)** UV-visible absorption spectra and room-temperature photoluminescence (PL) spectra of InGaN/GaN MQWs nanowires.

The optical properties of InGaN/GaN MQW NW PDs were studied through photoluminescence (PL) spectroscopy, and also UV-visible spectroscopy, as shown in Figure 1(c). The absorption edge is observed at around 386 nm and 350 nm, whereas the PL emission peak appears at approximately 511 nm. The red shift of the PL emission peak regarding to the absorption peak, reflecting a Stokes shift of approximately 1.12 eV, arises due to the strain-induced polarization effect upon 25% In incorporation.

The photoresponse of the planar and NW InGaN/GaN MQW was investigated using a standard three-electrode system with the formation of PEC-type photodetectors connected to the electrochemical workstation to measure the electrical properties of the device. In the three-electrode system, InGaN/GaN NWs device, Ag/AgCl and Pt electrodes were used as working electrode (WE), reference electrode (RE) and counter electrode (CE), respectively (Figure 2(a)). Besides, 1M NaOH aqueous solution is used as the electrolyte. Upon light illumination on the InGaN/GaN NWs, the electrons are excited from the valence band to the conduction band of the InGaN/GaN MQWs; the photogenerated electrons and holes are then separated due to the built-in electric field and participate in redox reactions. The upward surface band bending facilitates holes to migrate towards the NWs-electrolyte interface to initiate the oxidation reaction: $2H_2O + 4h^+ \rightarrow O_2 + 4H^+$. While the photogenerated electrons are transferred towards the counter (CE) electrode through the external circuit, participating in the reduction reaction: $2H^+ + 2e^- \rightarrow H_2$.

A comparative study of PEC-PD performance of planar and NW structure of InGaN/GaN MQWs devices is explored. Figure 2(b) and Figure. S1 shows the linear sweep voltammetry characteristics of NWs and planar devices with different light power densities of 365 nm light illumination and also under dark conditions. As the light power density rises with a small bias voltage (from negative to positive), the photocurrent density also increases for both devices. Higher light intensity generates a greater number of electron-hole pairs, and applying a positive bias voltage facilitates efficient carrier separation. As a result, the photogenerated holes drift towards NW-electrolyte interface, while the electrons accelerate towards the external circuit, leading to an increased photocurrent density. However, in comparison with the planar device, NW based photodetectors demonstrated higher photocurrent density. The maximum photocurrent density was obtained at around 16 mA/cm$^2$ under a light intensity of 43.5 mW/cm$^2$, which is almost double that of the planar device under self-powered conditions. The improved PEC performance of the NW photoanode is due to its higher surface-to-volume ratio, anti-reflection structures for the incident light, and the large solid–liquid interface that boosts active reaction sites.

To explore the spectral selectivity of the NW photodetector, we measured photocurrent density under different light irradiations (265 nm, 365 nm, 405 nm, and 530 nm) with a light power intensity of 5mW/cm$^2$ (Figure 2(c)). Our results show that the InGaN/GaN NW PD exhibits greater photoactive in the UV region, with maximum photocurrent density occurring at a wavelength of 365 nm. The photoresponse diminishes as it approaches the visible light wavelength and the UV-visible rejection ratio ($R_{365nm}/R_{530nm}$) was calculated to be 388.

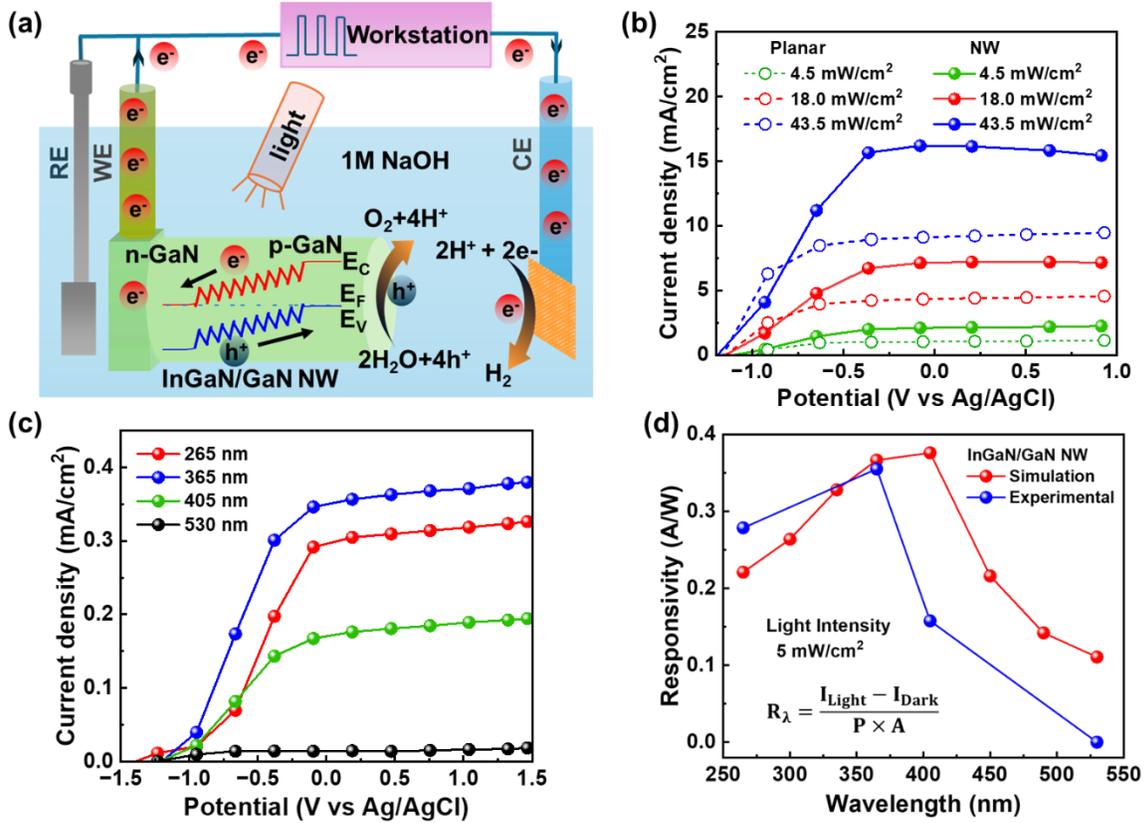

**Figure 2. (a)** Schematic of the fabricated NW MQW PEC-PD with three electrode PEC workstation. **(b)** Linear sweep voltammetry (LSV) characteristics of planar and NW based InGaN/GaN MQW device. **(c)** LSV under the irradiation of different light wavelength LEDs at light intensity 5 mW/cm². **(d)** Spectral photoresponse with both the simulated and experimental of the NW PEC photodetector device.

To assess the photodetection performance after photocurrent density, responsivity ($R_\lambda$) and detectivity ($D^*$) are two important performance metrics. Responsivity is determined by the photocurrent generated per unit incident light power per unit area and can be written as [9, 10]:

$$R_\lambda = \frac{I_{Light} - I_{Dark}}{P \times A} \quad \ldots\ldots\ldots\ldots\ldots\ldots (1)$$

where $I_{Light}$ and $I_{Dark}$ are the photocurrent density calculated under the dark and illumination condition, P and A are the illuminated light intensity and active device area, respectively. Detectivity ($D^*$), which is defined as the capability of detecting minimum intensity light signals and is calculated as [35]: 
$$D^* = \frac{R_\lambda}{\sqrt{2eI_{Dark}}} \quad \ldots\ldots\ldots\ldots\ldots\ldots (2)$$

where e is the electronic charge (1.6 x 10$^{-19}$C) and I$_{Dark}$ is the dark current density. A comparison study of responsivity with both simulated and experimental conditions with different spectral light is depicted in Figure 2(d). To calculate the photoresponse of the InGaN/GaN nanowires theoretically, finite difference time domain (FDTD) based numerical analysis was performed based on Maxwell's equations in three dimensions. The absorption profile obtained from the FDTD analysis was utilized to calculate the carrier generation rates in the nanowire array by integrating the number of absorbed photons per unit volume over the entire wavelength range of the illuminating light source. Subsequently, the following one-dimensional Poisson's equation and continuity equations were self-consistently solved in a coupled manner to obtain the electron and hole current densities, $J_n$ and $J_p$ Respectively.

$$\frac{\varepsilon_0 \varepsilon_r}{q} \frac{\partial^2 \phi(x)}{\partial x^2} = p(x) - n(x) + N_D^+(x) - N_A^-(x) - \sum_{trap} \rho_{trap}(x) \ldots \ldots \ldots \ldots (3)$$

$$\frac{\partial n(r,t)}{\partial t} = \frac{1}{q} \nabla \cdot J_n(r,t) + G_n(r,t) - R_n(r,t) \ldots \ldots \ldots \ldots (4)$$

$$\frac{\partial p(r,t)}{\partial t} = -\frac{1}{q} \nabla \cdot J_p(r,t) + G_p(r,t) - R_p(r,t) \ldots \ldots \ldots \ldots (5)$$

Here, in Equation (1), $\varepsilon_r$ represents the relative dielectric constant of the material, $\phi$ denotes the electric potential, *p(x)* and *n(x)* refer to the local densities of holes and electrons respectively, $N_A^-(x)$ is the ionized acceptor concentration, $N_D^+(x)$ is the ionized donor concentration, and $\rho_{trap}$(x) is the density of locally trapped charges. In Eqns. (2)-(3), the electron (hole) generation and recombination rates are represented by $G_n(G_p)$ and $R_n(R_p)$ respectively. More details of the numerical analysis are available in the Supplementary Information (SI) Table S1. As can be observed from Figure 2(d), both the experimental and theoretically calculated responsivity show a similar trend. The device shows a maximum responsivity peak at 0.33 A/W under the irradiation of 365nm wavelength with 5 mW/cm$^2$ power intensity. To study the self-powered photodetection properties, the current density-time characteristics were measured under the illumination of 365nm wavelength with different light intensities ranging from 0.7 mW/cm$^2$ to 43.5mW/cm$^2$ at zero applied bias (Figure S2). As light intensity increases, a greater number of electron-hole pairs are generated. These photogenerated carriers are subsequently separated and transported to the NW/electrolyte interface, increasing photocurrent.

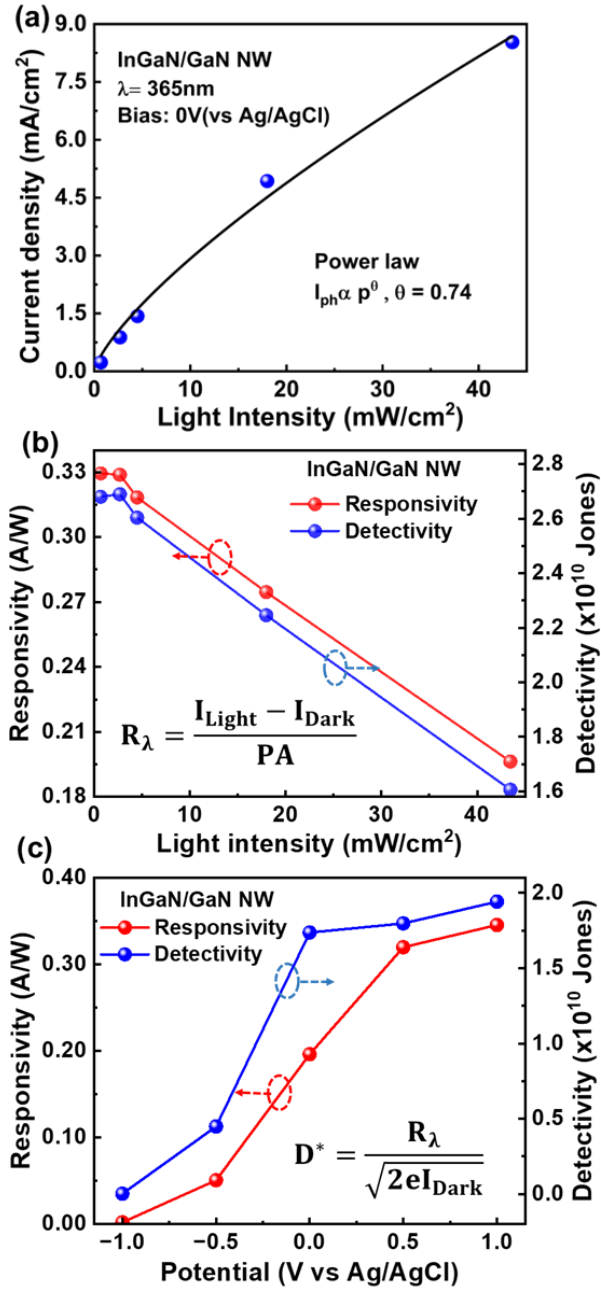

**Figure 3. (a)** Photocurrent density variation with illuminated light intensities fitted with the power law. **(b and c)** Responsivity and detectivity of the InGaN/GaN NW PEC photodetector device with different light

intensities under self-powered mode and external bias upon 365 nm light irradiation (Light intensity of 43.5 mW/cm$^2$).

Figure 3 (a) shows the rise in photocurrent density with the increase of the light power density. The photocurrent density is fitted using the power law (I α P$^θ$), and the θ is determined as 0.74. The high θ value indicates that effective charge carrier generation and separation happen; however, less than unity (ideal case) suggests that recombination and charge carrier trapping contribute to the photocurrent generation process.

The device responsivity and detectivity as a function of the light power density upon 365 nm irradiated light at zero applied bias voltage are shown in Figure 3(b). The device exhibits excellent responsivity and detectivity of 0.33 A/W and 2.67 x 10$^{10}$ Jones, respectively, under the irradiation light intensity of 0.7 mW/cm$^2$ in self-powered mode, indicating that the device has a strong response ability at a lower intensity of UV light. The nanowire geometry of MQW and the generated built-in electric field allow for efficient charge carrier separation and transmission. Both the responsivity and detectivity decrease with the increase of light intensity. This is attributed to scattering and recombination of charge carriers at higher light intensities. Furthermore, bias-dependent responsivity and detectivity are calculated and displayed in Figure 3(c). Both parameters increase with increasing applied forward bias. The device shows maximum responsivity and detectivity of 0.34 A/W and 1.94 x 10$^{10}$ Jones at an applied voltage of 1.0 V vs Ag/AgCl under the irradiated light intensity of 43.5 mW/cm$^2$. The increase and maxima values of responsivity and detectivity are mainly due to the voltage-dependent charge carrier transfer from the electrode to the redox couple, which improves photocurrent density by reducing the interfacial energy barrier and enhancing charge injection efficiency.

We have also studied the photoresponse of the PEC photodetection at different applied biases and various illumination intensities with an excitation wavelength of 365 nm (Figure 4(a)). The photocurrent density rises with an increase in the forward applied bias; however, it decreases with the reverse bias. In the forward bias, the InGaN/GaN MQWs band bends downward at the semiconductor/electrolyte interface, which facilitates the photo-generated electron-hole pair separation and their transportation in the opposite direction. Therefore, the device shows enhanced photo-response under positive bias. However, under reverse bias, the bands gradually become flat, which leads to charge carrier recombination and inhibits the photocurrent.

To get further insights into the photoresponse characteristics of NW PD, the response time ($t_r$) and recovery time ($t_d$) are also measured. Response/recovery time is defined as the time taken by the photocurrent to reach/decay from its 10/90% to 90/10% value. As depicted in Figure 4 (b), under 365 nm light illumination with light intensity of 43.5 mW/cm², the response and recovery time of the NW PEC photodetector device are estimated as 168 ms and 69 ms respectively without external source of energy. The fast detection speed validates the capability of the device to detect rapidly changing optical signals.

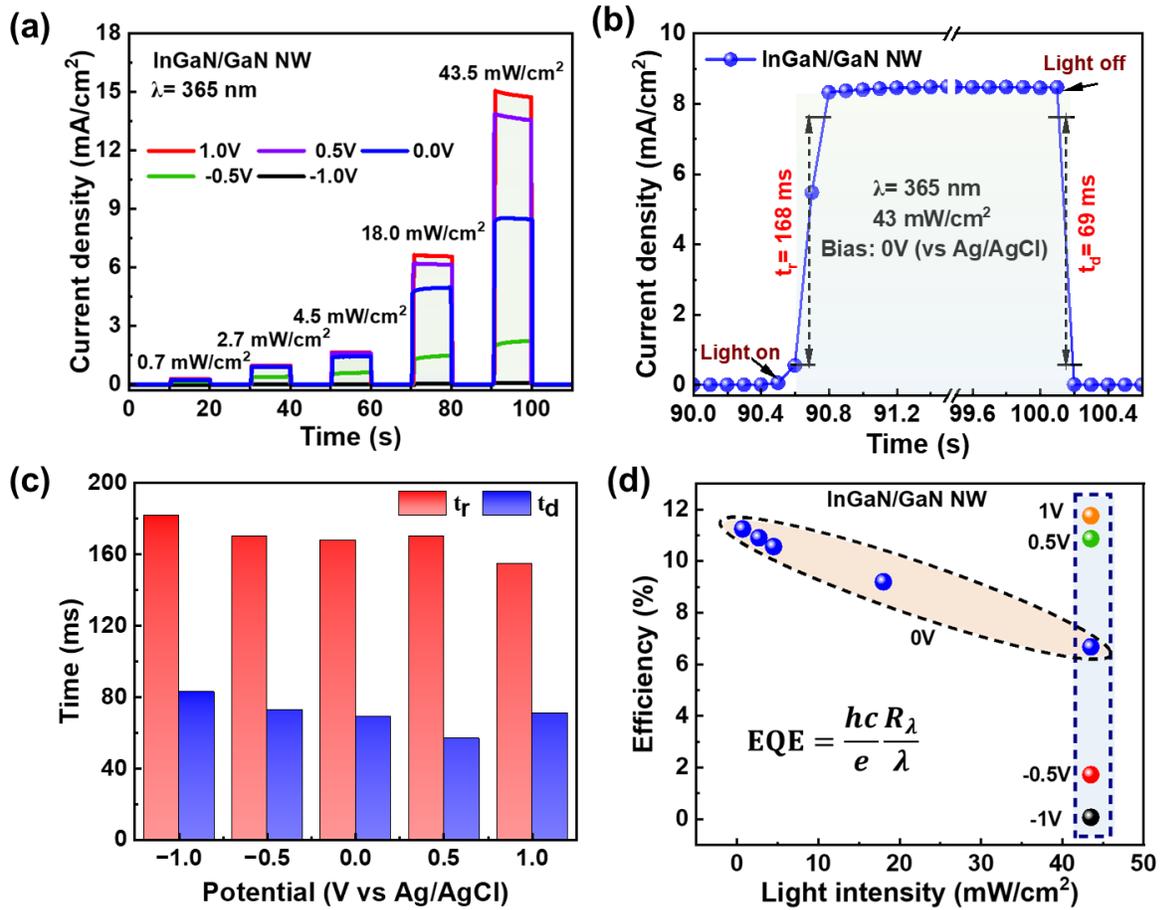

**Figure 4. (a)** Transient photocurrent density of InGaN/GaN MQW NW based PD device under external applied bias voltage and different light intensities. The photo response of the device can be tuned by applied bias and irradiated light intensities. **(b)** Photocurrent response under light on/off switching modes at zero applied bias voltage. **(c)** Rise and recovery time under 365 nm light illumination with light intensity of 43.5 mW/cm² at different bias conditions. **(d)** Calculated EQE under different incident optical power intensity and external applied bias.

The influence of the external bias voltage on the photoresponse of NW photodetector is explored with a light intensity of 43.5 mW/cm² (Figure 4 (c)) and (Figure S3). The calculated $t_r$ and $t_d$ time values gradually decrease with the increase of the bias voltage, and finally $t_r$ reach a minimum value of 148 ms at 1 V vs Ag/AgCl. The gradual decrease of both rise and recovery time values indicates that charge carriers are affected by the environment.

To study the photosensitivity of the NW photodetector, we studied the external quantum efficiency (EQE) which value is related to the absorption coefficient of the device, and the separation and transport efficiency of the photogenerated charge carriers. The EQE can be calculated from the following equation [36]:

$$\text{EQE} = \frac{hc}{e}\frac{R_\lambda}{\lambda} \dots\dots\dots\dots\dots\dots. (6)$$

Where $\lambda$ is the incident light wavelength, $c$ is the speed, $h$ is the Planck's constant. Figure 4(d) displays the influence of the light intensity and the external bias voltage on the EQE. As can be seen, at 0V vs Ag/AgCl, the EQE decreases from around 11% to around 7% with the increase of the light intensity from 0.7 mW/cm² to 43.5 mW/cm². Also, the EQE values were calculated with different external bias voltages under a stable light intensity of 43.5 mW/cm². The EQE value increased from 7% at 0 V vs Ag/AgCl to 11% at 1 V vs Ag/AgCl, indicating that the EQE is mainly affected by the interaction of the light intensity and the external bias energy sources. Therefore, according to the light intensity, the external bias can be determined for obtaining the desired EQE value.

Finally, the self-powered photoresponsivity of InGaN/GaN NW with different light wavelength in this work and other reported PEC photodectors are displayed in Figure 5. The NW PD in this study shows better responsivity at 365 nm, which is higher than the value of the III- nitride based PEC-PDs reported yet.

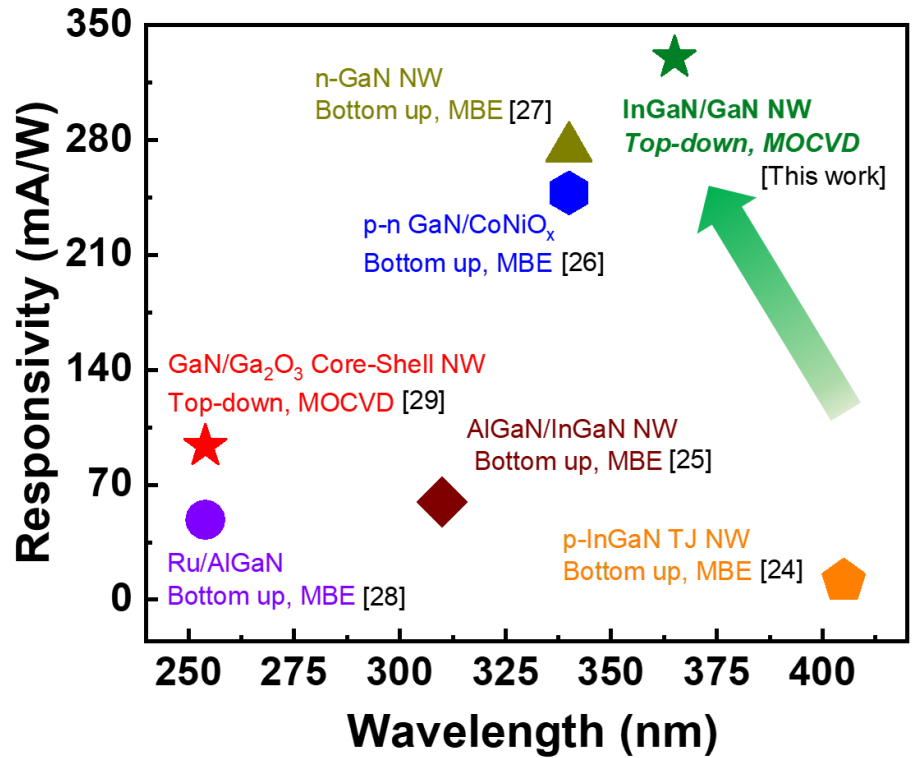

**Figure 5.** Comparison of the responsivity of the InGaN/GaN PEC-PD with other III- nitride based PEC-PDs grown or fabricated by different techniques (MBE/MOCVD growth technique or bottom-up/top-down fabrication technique) and measured under light sources of different wavelengths.

**Conclusion**

In conclusion, a UV-visible, self-powered NW-based InGaN/GaN MQW heterojunction PEC-PD have been explored. Uniform InGaN/GaN MQW NWs were fabricated using the silica nanosphere lithography technique coupled with the top-down approach. The PEC PD with NW structures showed higher photoresponse under self-powered mode than from planar photodetector. The enhanced photoresponse in NW photodetector is mainly due to the efficient charge carrier separation and hole migration to the NW surface due to the built-in electric field. Moreover, the device demonstrates self-powered UV light PEC photodetection characteristics, and performance can be extended towards the visible region. The device exhibits maximum responsivity and detectivity of 0.33 A/W and 2.67 x $10^{10}$ Jones, respectively, under the irradiated light intensity of 0.7 mW/cm$^2$ and wavelength of 365 nm, which is highest among the reported III-nitride based PEC photodetectors. Also, we have examined the bias-dependent photo response of the device. It turns out that the photo response is enhanced at positive applied bias whereas is reduced at reverse bias. Therefore, the photo response can be tuned by the external applied bias. These results validate that MQW hetero-structures and nanostructuring then can be for a scalable and viable platform for high-performance PEC photodetection.

**Experimental Section**

**Fabrication of Nanowires**: The MOCVD grown InGaN/GaN MQW NW photodetector was fabricated using silica nanosphere patterning technique. Firstly, commercially available silica nanospheres of 365 nm diameter in aqueous solution (Microparticles, GmbH) were uniformly coated on the top surface p-GaN of the MQW. The nanowires were fabricated using two steps. First, silica nanosphere (diameter - 365 nm) was deposited as a hard mask on the thin film samples using spin coating method. After that, the coated sample is etched by $Cl_2$/Ar based inductively coupled reactive ion etching (ICP-RIE) for the formation of NW. The NWs were then treated with KOH based buffer solution to passivate the plasma-damaged sidewalls due to dry etching. Finally, silica nanospheres were removed from the etched nanowires using sonication with DI water. The average height and diameter of the fabricated NW were around 900 nm and 330 nm, with a surface coverage of 50%.

**Structural and optical characterizations:** The morphology of the fabricated InGaN/GaN MQWs NW based photodetector device was examined by the scanning electron microscope (SEM, TESCAN, VEGA 3). The absorption properties were analyzed using UV-visible spectrometer (Lambda 750, PerkinElmer) at near normal incidence ($\theta = 8°$). The photoluminescence (PL) studies were carried out using home built μ-PL setup. A continuous wave laser diode (wavelength of 405nm) were used as excitation source.

**PEC-Photodetection measurements**: The photoelectrochemical characterizations of InGaN/GaN MQWs based photodetector devices were investigated using three three-electrode set up in an elcterochemical cell with quartz window. The device is used as a working electrode whereas platinum and Ag/AgCl electrode are used as counter and reference electrode, respectively. 1M NaOH aqueous solution (pH~12) was employed as electrolyte in the PEC photodetection measurements. Before the measurement, a contact window was patterned by standard lithography technique on the MQW to etch it down to the n-GaN film. Later, an ohmic contact was made on the n-GaN film by applying Ga-In eutectic (99.99%, Sigma Aldrich) coupled with the silver paste (Ted Pella) and copper tape. Finally, the sample was encapsulated with an insulating epoxy to prevent the leakage current. Electrochemical measurements were carried out using a potentiostat (1010E, Gamry). Different wavelength LEDs (265 nm, 365 nm, 405 nm and 532 nm, Thorlabs mounted LEDs) were used to irradiate the device and the optical light intensity was measured using a calibrated Thorlabs S120VC photodiode.

**Acknowledgements:**

This work was supported by the Natural Sciences and Engineering Research Council of Canada (NSERC) through Discovery and Alliance Grant programs, the MEIE Photonique Quantique Quebec (PQ2) program, Fonds de Recherche du Québec - Nature et Technologies (FRQNT) and the Canada Research Chair program. We would like to acknowledge CMC Microsystems for the provision of products and services and fabrication fund assistance using the facilities of Laboratory of Mico and Nanofabrication (LMN) at INRS.

**Supplementary information**

**Experimental Results:**

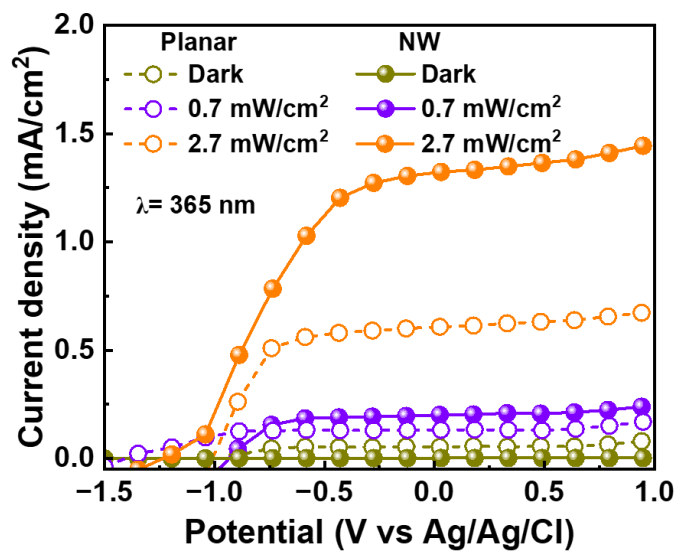

**Figure S1:** Linear sweep voltammetry (LSV) characteristics of planar and NW based InGaN/GaN MQW device under different light intensities and also under dark condition.

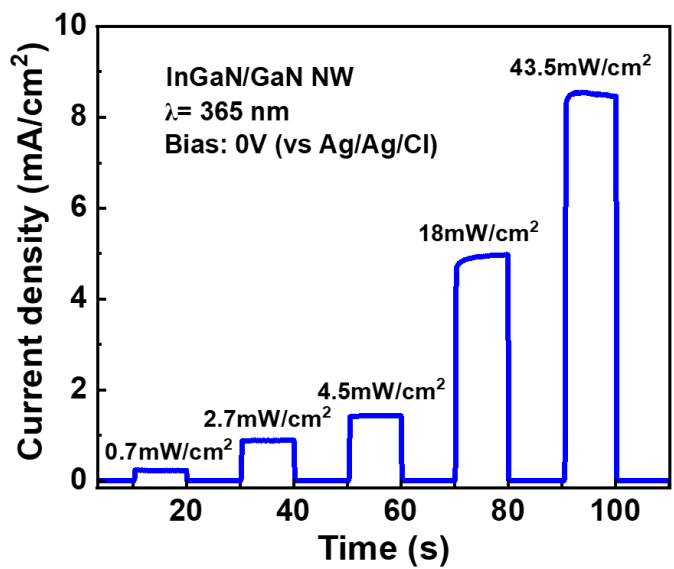

**Figure S2:** Transient photocurrent density of InGaN/GaN MQW NW based PD device at self-powered condition and different light intensities.

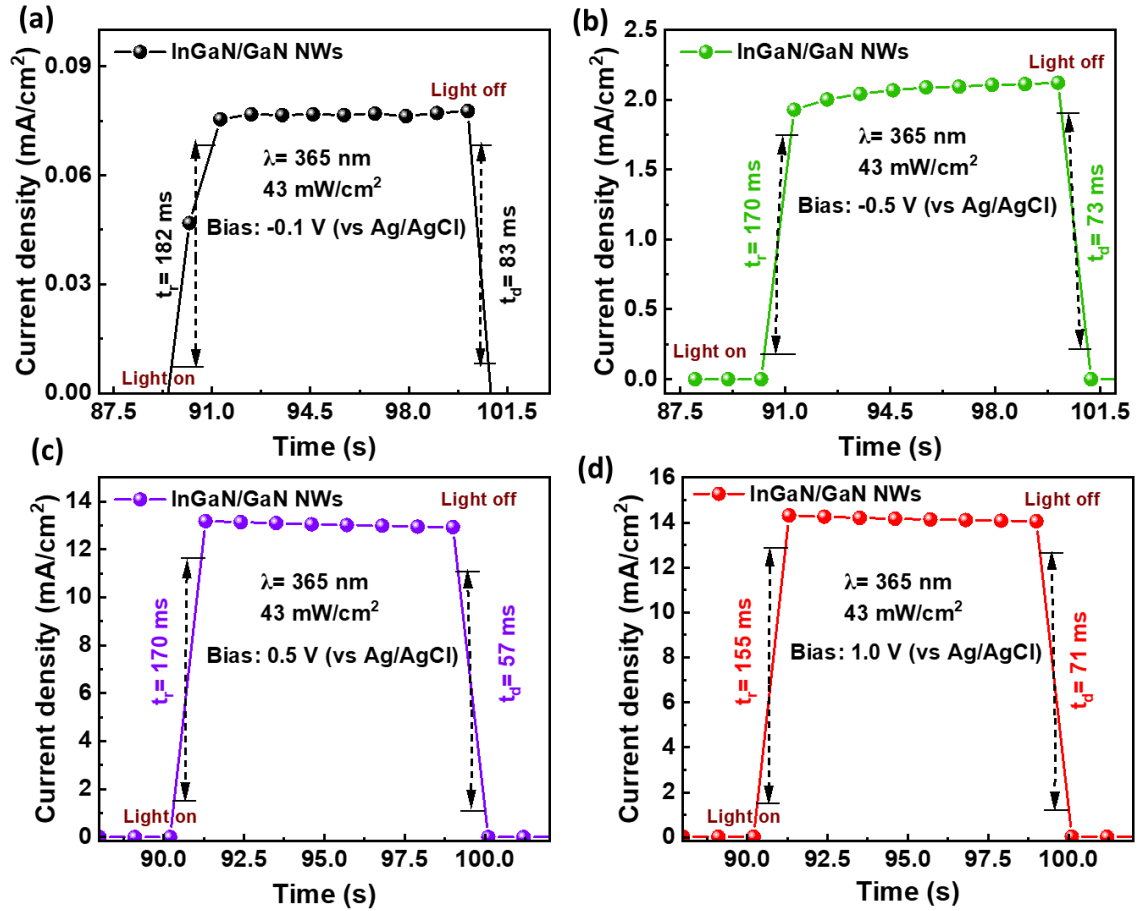

**Figure S3:** Photocurrent response under light on/off switching modes at different applied forward and reverse bias voltage.

**Theoretical Analysis:**

To obtain the absorption characteristics of the nanowire arrays, three-dimensional finite-difference time-domain (FDTD) simulations were performed using Ansys Lumerical suite. Representative InGaN/GaN nanowire arrays having finite heights of 800 nm and volumetric fill factor of 49.8% were considered in the FDTD simulation. Illuminating light sources, having peak emission wavelengths ranging from 250 nm to 550 nm and spectral bandwidth of 50 nm, have been considered to calculate the absorption characteristics of the nanowire arrays. The carrier generation rate is subsequently calculated by integrating the number of absorbed photons per unit volume over the entire wavelength range of the illuminating spectrum.

The spatial variation of electric potential caused by electric charges within the photodetector is be determined by solving Poisson's equation.

$$\frac{\varepsilon_0 \varepsilon_r}{q} \frac{\partial^2 \phi(x)}{\partial x^2} = p(x) - n(x) + N_D^+(x) - N_A^-(x) - \sum_{trap} \rho_{trap}(x) \quad \ldots \ldots \ldots \ldots (1)$$

Here, $\varepsilon_r$ represents the relative dielectric constant of the semiconductor layer, $\phi$ denotes the electric potential, p(x) and n(x) refer to the local densities of holes and electrons respectively, $N_A^-(x)$ is the ionized acceptor concentration, $N_D^+(x)$ is the ionized donor concentration, and $\rho_{trap}$(x) indicates the density of locally trapped charges. The spatial variation of potential is used to evaluate the drift and diffusion currents based on the following relations:

$$J_{drift} = qE(n\mu_n + p\mu_p) \quad \ldots \ldots \ldots \ldots (2)$$

$$J_{nx|diff} = qD_n \nabla n \quad \ldots \ldots \ldots \ldots (3)$$

$$J_{px|diff} = -qD_p \nabla p \quad \ldots \ldots \ldots \ldots (4)$$

Here $n$ ($p$) is the total electron (hole) density, $\mu_n(\mu_p)$ is the electron (hole) mobility, $D_n$ ($D_p$) is the electron (hole) diffusion coefficient, and $\nabla n$ ($\nabla p$) is the electron (hole) concentration gradient. The electric field $E$ is related to the electric potential ($\phi$) according to the relation $E = -\nabla \phi$. The total hole and electron current density is obtained from the drift and diffusion current densities as follows.

$$J_n = qn\mu_n E + qD_n \nabla n \quad \ldots \ldots \ldots \ldots (5)$$

$$J_p = qp\mu_p E - qD_p \nabla p \quad \ldots \ldots \ldots \ldots (6)$$

Subsequently, the current density and generation profile derived from the FDTD analysis and Eqns. (1)-(6) are incorporated into the continuity equations, which are defined as follows:

$$\frac{\partial n(r,t)}{\partial t} = \frac{1}{q} \nabla \cdot J_n(r,t) + G_n(r,t) - R_n(r,t) \quad \ldots \ldots \ldots \ldots (7)$$

$$\frac{\partial p(r,t)}{\partial t} = -\frac{1}{q} \nabla \cdot J_p(r,t) + G_p(r,t) - R_p(r,t) \quad \ldots \ldots \ldots \ldots (8)$$

Here, $G_n(G_p)$ and $R_n(R_p)$ represent the electron (hole) generation and recombination rates respectively. To calculate recombination rates, radiative band to band recombination rate ($R^R_{n,p}$), Auger recombination rate ($R^A_{n,p}$) and Shockley-Read-Hall recombination rate ($R^{SRH}_{n,p}$ (r,t)) have been considered. Total recombination rate is given by:

$$R_{n,p} = R^R_{n,p} + R^A_{n,p} + R^{SRH}_{n,p} \quad\ldots\ldots\ldots\ldots (9)$$

A list of different material parameters considered in the simulation and the corresponding references are listed in Table S1.

Table S1: List of material parameters considered in the theoretical analysis

| Parameter Name | Value |
|---|---|
| ITO Work Function [1] | 4.5 eV |
| GaN electron mobility [2] | 900 cm$^2$/V-s |
| GaN hole mobility [3] | 10 cm$^2$/V-s |
| GaN radiative recombination coefficient [4] | 1.5x10$^{-11}$ cm$^3$/s |
| GaN Auger recombination coefficient for electrons (holes) [5] | 1.4x10$^{30}$ cm$^6$/s (1.4x10$^{30}$ cm$^6$/s) |
| InN relative dielectric constant [6] | 15.3 |
| InN Work Function [7] | 5.2 eV |
| InN electron effective mass, m$_n$ [8] | 0.11m$_0$ |
| InN electron mobility [8] | 1100 cm$^2$/V-s |
| InN hole mobility [9] | 39 cm$^2$/V-s |
| InN Radiative Recombination Coefficient [10] | 5.2x10$^{11}$ cm$^3$/s |
| InN Auger Recombination coefficient for electrons (holes) [10] | 1.2 x10$^{28}$ cm$^6$/s (1.2 x10$^{28}$ cm$^6$/s) |
| AlN relative dielectric constant [11] | 8.5 |
| AlN electron effective mass [12] | 0.4m$_0$ |
| AlN hole effective mass [11] | 3.53m$_0$ |
| AlN electron (hole) mobility [11] | 300 cm$^2$/V-s (14 cm$^2$/V-s) |
| AlN Work Function [14] | 5.35 eV |
| AlN Radiative Recombination Coefficient [13] | 1.68 x10$^{11}$ cm$^3$/s |
| AlGaN Auger Recombination Coefficient for electrons (holes) [15] | 5 x10$^{32}$ cm$^6$/s (5 x10$^{32}$ cm$^6$/s) |